\RequirePackage{fix-cm}
\documentclass[twocolumn,epjc3]{svjour3}  
\smartqed  
\RequirePackage{graphicx}
\usepackage{multirow}
\usepackage{amsmath}
\usepackage{xcolor}
\usepackage{lineno}
\usepackage{amssymb}
\usepackage{upgreek}
\journalname{Eur. Phys. J. C}
\begin{document}

\title{Radon daughter removal from PTFE surfaces and its application in liquid xenon detectors
}


\author{S. Bruenner\thanksref{e1,e3,addr1}
        \and
        D. Cichon\thanksref{addr1} 
        \and
        G. Eurin\thanksref{e4,addr1}
        \and
        P. Herrero G\'omez\thanksref{e5,addr1}
        \and 
        F. J\"org\thanksref{addr1}
        \and
        T. Marrod\'an Undagoitia\thanksref{addr1}
        \and
        H. Simgen\thanksref{addr1}
        \and
        N. Rupp\thanksref{e2,addr1}
}

\thankstext{e1}{e-mail: stefanb@nikhef.nl}
\thankstext{e2}{e-mail: natascha.rupp@mpi-hd.mpg.de}
\thankstext{e3}{now at: Nikhef, Science Park, 1098XG Amsterdam, Netherlands}
\thankstext{e4}{now at: CEA/Saclay, IRFU (Institut de Recherche sur les Lois Fondamentales de l’Univers), F-91191 Gif-sur-Yvette CEDEX, France}
\thankstext{e5}{now at: Donostia International Physics Center, 20018 Donostia - San Sebasti\'an, Spain}


\institute{Max-Planck-Institut f\"ur Kernphysik, Heidelberg, D-69117 Heidelberg, Germany \label{addr1}
}

\date{Received: date / Accepted: date}

\maketitle

\begin{abstract}
Long-lived radon daughters are a critical background source in experiments searching for low-energy rare events. Originating from radon in ambient air, radioactive polonium, bismuth and lead isotopes plate-out on materials that are later employed in the experiment. In this paper, we examine cleaning procedures for their capability to remove radon daughters from PTFE surfaces, a material often used in liquid xenon TPCs. We found a large difference between the removal efficiency obtained for the decay chains of $^{222}$Rn and $^{220}$Rn, respectively. This indicates that the plate-out mechanism has an effect on the cleaning success. While the long-lived $^{222}$Rn daughters could be reduced by a factor of ~2, the removal of $^{220}$Rn daughters was up to 10 times more efficient depending on the treatment. Furthermore, the impact of a nitric acid based PTFE cleaning on the liquid xenon purity is investigated in a small-scale liquid xenon TPC. 
\end{abstract}

\section{Introduction}
\label{sec:intro}
In low-energy experiments searching for rare signals, such as dark matter interactions or neutrinoless double beta decay \cite{strigari,neutrino}, it is of utmost importance to suppress backgrounds to a negligible level. Radon progenies can plate-out on the detector surface when exposing it to air, which contains radon. An important background contribution arises from the $^{222}$Rn decay products subsequent to the long-lived $^{210}$Pb (T$_{1/2}=$ 22.3 years), as they persist during the entire run-time of such experiments  \cite{superCDMS,counterplateout,borexinoplateout,SR1paper,LUXfullexposure}.\\ 
PTFE (polytetrafluoroethylene) is commonly used in many low-background experiments and often large surface areas of this material are exposed to the active volume of the detector \cite{pandaX,LZ,xenon_instrument,darwin,nexo,cuore,gerda}. It is a material that tends to carry negative charges, due to the triboelectric effect \cite{te_zou,te_henniker,te_liu}, whereas radon daughters have a high probability to be positively charged \cite{ion_fraction,exo_ion}. Therefore, their plate-out probability onto PTFE surfaces is enhanced in comparison to e.g. steel surfaces \cite{morrison:plateout_teflon}. Background events may be induced by direct radiation from $\alpha$-, $\beta$- and $\gamma$-decays. Moreover, $\alpha$-decays of $^{210}$Po may induce background by the recoiling nucleus $^{206}$Pb as well as by neutrons. The latter are generated by ($\alpha$,n)-reactions on the fluorine in PTFE \cite{neutron_ptfe}.
\begin{figure*}[htbp]
\centering 
\includegraphics[width=0.75\textwidth]{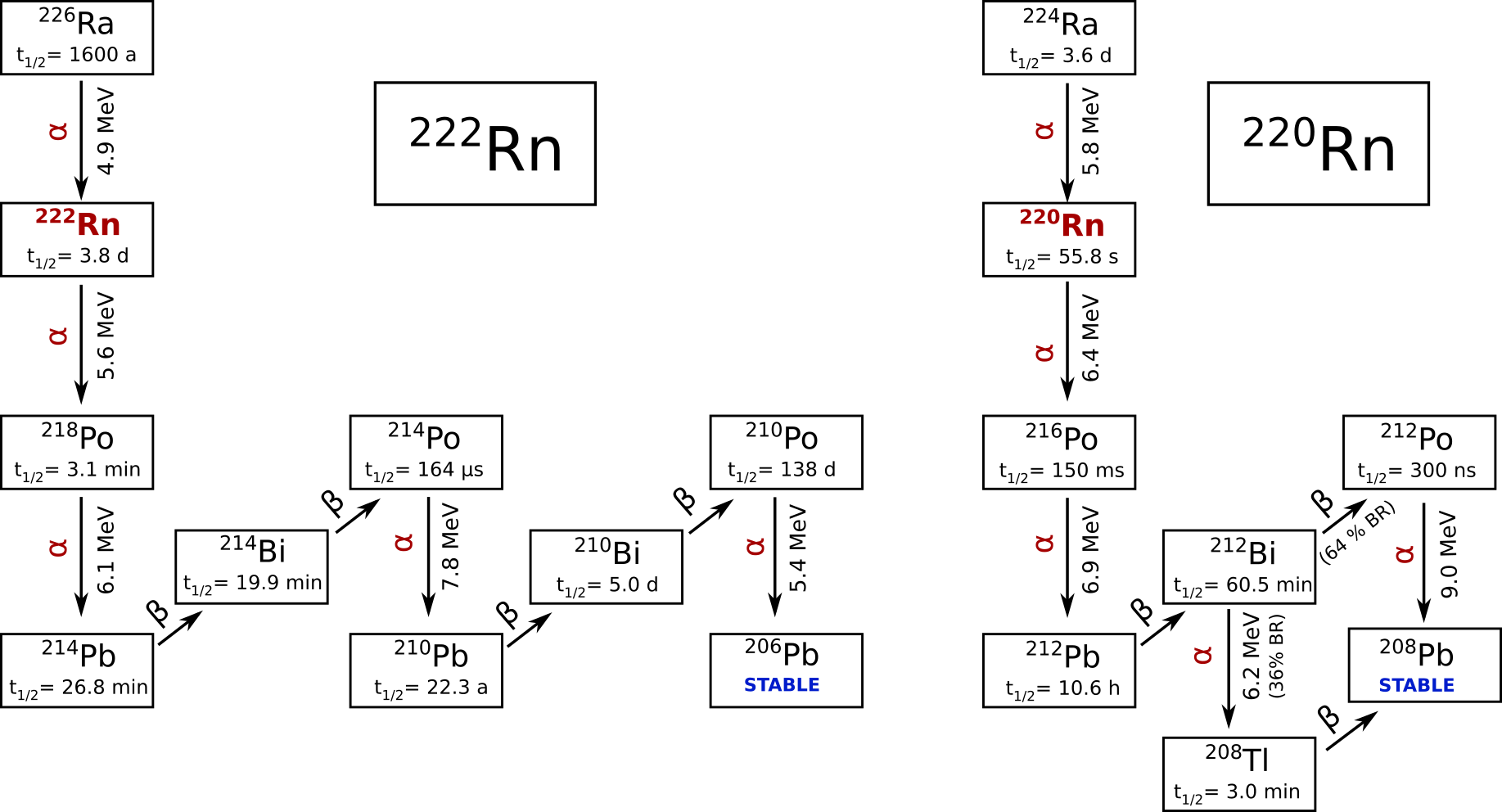}
\caption{\label{fig:radon_chain} Decay chains of $^{222}$Rn (left) and $^{220}$Rn (right), respectively. While all $^{220}$Rn daughters are short-lived, the $^{222}$Rn chain is typically broken due to the long-lived lead isotope $^{210}$Pb.}
\label{fig:decay-chain}
\end{figure*}
\\
In this work, we present different treatments for 
PTFE surfaces for the removal of plated-out decay daughters originating from the $^{222}$Rn and $^{220}$Rn decay chains. Both decay chains, including half-life and decay energies, are shown in Fig. \ref{fig:decay-chain}.
A commonly used detector type for rare-event searches are noble gas Time Projection Chambers (TPCs). To employ a PTFE surface treatment in such a detector, it has to be validated that the detector's signal production is not impacted by chemical residuals from the cleaning. 
\\
In section \ref{sec:1.1}, the preparation of PTFE samples loaded with radon daughters and the measurement of their surface activity is described. In section \ref{sec:1.2}, surface cleaning procedures for the removal of $^{222}$Rn daughters are presented. As the $\beta$-decay of lead is a central source of background in dark matter experiments, we performed a more thorough study on its removal in section \ref{sec:1.3}, employing a $^{220}$Rn source. In section \ref{sec:1.4}, the results on the lead removal from both radon chains are compared. In section \ref{2}, we test the application of a selected cleaning method in a local liquid xenon TPC. 

\section{Surface treatments for radon daughter removal}
\label{sec:1}

\subsection{Sample preparation and measurement of surface activity}
\label{sec:1.1}
\begin{figure*}[htbp]
\centering 
\includegraphics[width=1.0\textwidth]{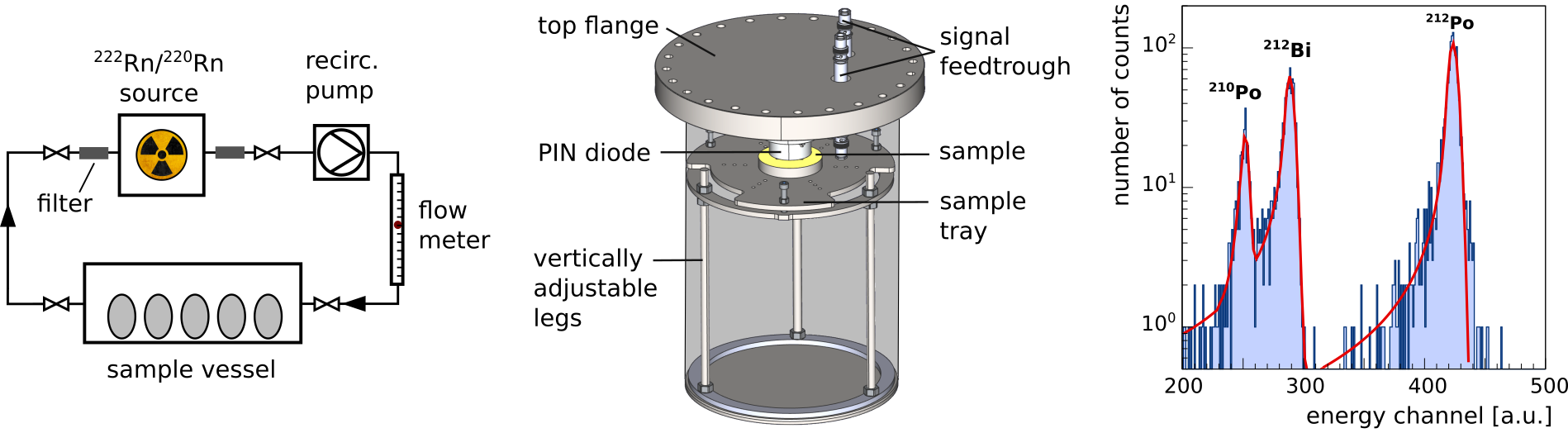}
\caption{\label{fig:exp_setup} (Left) Schematic drawing of the loading set-up. The PTFE samples were placed in a vessel which was flushed with radon enriched air from an emanation source. (Middle) The surface activity of the loaded samples was measured by the depicted alpha-spectrometer. During measurement, the spectrometer was under vacuum. An adjustable sample tray allows to place the sample close to the sensitive silicon PIN photo diode. (Right) Exemplary alpha-spectrum as it was obtained when measuring $^{220}$Rn daughters after cleaning. }
\end{figure*}

For our study we used circular $0.5\,$mm thick PTFE plates with a diameter of $50\,$mm. In order to enrich the samples surfaces with radon daughters, we employed the set-up sketched in Fig. \ref{fig:exp_setup} (left). The sample discs were placed in a vessel which is connected to a $^{222}$Rn or $^{220}$Rn emanation source, depending on the radon chain under investigation. For the plate-out of $^{222}$Rn daughters, we used uranium oxide powder as a source which emanates $^{222}$Rn at a rate of $\sim 1\,$MBq. A recirculation pump ensured a constant gas flow of $\sim 1 \,$slpm of radon enriched air through the sample vessel. 
The loading of $^{222}$Rn daughters was done over a time period of two years, including short interruptions for maintenance of the set-up. Thereafter, the samples were stored for another two years prior to our measurements without any further loading. Consequently, we can assume in good approximation that the long-lived radon daughters $^{210}$Pb, $^{210}$Bi and $^{210}$Po are in secular equilibrium and independent from systematic irregularities during the loading process, such as changed or interrupted recirculation flows or different plate-out rates among the samples. For studying $^{220}$Rn daughters, we employed a $4\,$kBq $^{228}$Th source of a similar type as described in \cite{rn220_source}. The loading with non stable $^{220}$Rn daughters on the PTFE plates saturates already after a few days due to the relatively short half-lives in the decay chain (see Fig. \ref{fig:decay-chain}). For the loaded samples, we can assume secular equilibrium between $^{212}$Pb, which has the longest half-life of $10.5$\,h, and the subsequent daughter isotopes in the decay chain.\\

The surface activity of the loaded samples was measured by means of an $\alpha$-spectrometer which is sketched in Fig. \ref{fig:exp_setup} (middle). During the measurement, the sample is placed in the center of the sample tray facing a windowless silicon PIN-diode (Hamamatsu Si PIN photodiode model S3204-09) which is mounted on the top flange. The diode has a rectangular shape and $1.7$\,cm$^2$ of sensitive surface. The sample tray can be adjusted vertically in order to minimize the gap between diode and sample and thus optimize for the detection efficiency concerning the $\alpha$-emission's solid angle. For this study, the distance to the diode was kept constant at $6.1\,$mm which results in a detection efficiency of $0.32$ for the sample's side facing the diode \cite{pablo}. During the measurements, the $\alpha$-spectrometer was continuously pumped to avoid energy losses of the $\alpha$-particles in residual ambient gas. The anode and cathode signals from the diode are amplified, subtracted from each other and processed by a multi-channel analyzer (MCA). In the right panel of Fig. \ref{fig:exp_setup}, a typical energy spectrum is shown as it was acquired when studying the removal of $^{220}$Rn daughters. The peaks of $^{212}$Po and $^{212}$Bi are clearly visible. The latter is overlapping with the $^{210}$Po peak which is also present in the $^{220}$Rn data as a background. It is originating from the contamination of the spectrometer and the PIN-diode themselves due to an earlier exposure of the detector to $^{222}$Rn. During our studies, the $^{210}$Po background was monitored and taken into account when determining the $^{210}$Po reduction after the investigated procedures. The alpha peaks are fitted using the crystal-ball function (red line) as in \cite{crystal_ball}. It consists of a gaussian part and a left-sided tail which accounts for the asymmetric peak shape. In order to determine the activity, we chose a selection window of $+3\,\sigma$ and $-5\,\sigma$ around the fitted gaussian mean of the $^{210}$Po and $^{212}$Bi peak, respectively. For the well separated $^{212}$Po peak, the selection window was increased to $+5\,\sigma$ and $-10\,\sigma$. It should be noted that the $^{220}$Rn daughters were not present during our study of the $^{210}$Po removal. Since only relative activity changes are used to determine the reduction of radon daughters, we did not correct for the fraction of counts outside the selection window.


\subsection{Removal of $^{222}$Rn daughters from PTFE surfaces}
\label{sec:1.2}
In this study, we monitored the $^{210}$Po surface activity of four different PTFE discs loaded with long lived $^{222}$Rn daughters. After the initial activity was measured, we applied different cleaning procedures in sequence while monitoring the $^{210}$Po reduction after each cleaning step. In Table \ref{tab:c_proc} the investigated procedures are summarized.


\begin{table*}[t]
\centering
\renewcommand{\arraystretch}{1.2}
\smallskip
\begin{tabular}{c|l}
\textbf{cleaning procedure} & \textbf{description}\\
\hline
weak HNO$_3$ & 5\% nitric acid solution for 2\,h, including 15\,min in ultrasound\\ \hline
strong HNO$_3$ & 32\% nitric acid solution for 2\,h, including 15\,min in ultrasound\\ \hline
hot HNO$_3$ & 32\% nitric acid solution at 60$^\circ$C for 2\,h, including 15\,min in ultrasound\\ \hline
multiple HNO$_3$ & 15\,min 32\% nitric acid solution in ultrasound + \\
 & immerse sample three times in fresh nitric acid baths (32\%) for 90\,s each\\ \hline
sod. hydro. & 5\% sodium hydroxide for 2\,h, including 15\,min in ultrasound\\ \hline
ethanol wiping & wipe sample using a clean-room tissue soaked with ethanol \\ \hline
ethanol bath & ethanol bath for 15\,min while applying ultrasound\\ \hline
\end{tabular}
\caption{\label{tab:c_proc} Investigated cleaning procedures applied to PTFE samples loaded with $^{222}$Rn daughters. The acid concentrations are given by weight.}
\end{table*}
Nitric acid (HNO$_3$) is commonly used for the cleaning of PTFE parts \cite{cuore,exo_det_const}. In our study, we tested different nitric acid solutions of 5\% and 32\% concentration (by mass), respectively, and at room temperature. Another nitric acid cleaning was performed at an increased temperature of $60\,^{\circ}$C. During all those treatments, the PTFE sample was placed in 100\,ml of a nitric acid solution for 2\,h. For the first 15\,min, the glass containing the sample in the acid solution was immersed in an ultrasonic bath. After the treatment, the sample was thoroughly rinsed with DI-water. In case of the so-called multiple HNO$_3$ procedure, the PTFE was kept for 15\,min in a 32\% nitric acid solution while being simultaneously exposed to ultrasound. Then, the sample was immersed three times in a nitric acid solution of the same concentration for 90\,s each. Before every sample immersion, the solution was exchanged for a fresh one. This procedure was chosen in order to probe the potential re-deposition of $^{210}$Po on the surface similar to what was observed for copper \cite{zuzel_wojcik}. Do to the quick exchange of the acid, the exposure time is kept short such that re-deposition will be significantly reduced with respect to the other nitric acid procedures.\\
Besides procedures based on nitric acid, we also tested a sodium hydroxide treatment where the PTFE sample was immersed for 2\,h in the cleaning bath including 15\,min of ultrasound exposure. For the ethanol wipe procedure, the sample was thoroughly wiped using a clean-room tissue soaked with ethanol. Furthermore, we studied the effect of complete immersion of the sample into an ethanol bath including 15\,min of ultrasound treatment.
\begin{table}[h]
\centering
\renewcommand{\arraystretch}{1.0}
\smallskip
\begin{tabular}{r|c|c|c}
 & \textbf{disc 1} & \textbf{disc 2a}& \textbf{disc 2b}\\
 \hline
 start act. & $18.34 \pm 0.26$ & $22.4 \pm 0.8 $ & $7.75 \pm 0.22$ \\ \hline
 \multirow{2}{*}{1st step} & ethanol wiping & \multicolumn{2}{c}{ethanol wiping} \\
 & $11.20\pm0.04$ & $13.3\pm0.3$ & $3.99\pm0.12$\\ \hline
  \multirow{2}{*}{2nd step} & weak HNO$_3$ & \multicolumn{2}{c}{ethanol wiping} \\
 & $8.55\pm0.07$ & $10.68\pm0.20$ & $3.52\pm0.07$\\ \hline
   \multirow{2}{*}{3rd step} & strong HNO$_3$ & \multicolumn{2}{c}{strong HNO$_3$} \\
 & $7.64\pm0.02$ & $10.08\pm0.25$ & $3.22\pm0.07$\\ \hline
   \multirow{2}{*}{4th step} & hot HNO$_3$ & \multicolumn{2}{c}{multiple HNO$_3$} \\
 & $7.27\pm0.04$ & $8.8\pm0.5$ & $3.01\pm0.06$\\ \hline
   \multirow{2}{*}{5th step} & \multirow{2}{*}{-} & \multicolumn{2}{c}{sodium hydroxide} \\
 &  & - & $2.86\pm0.13$\\ \hline
   \multirow{2}{*}{6th step} & \multirow{2}{*}{-} & \multicolumn{2}{c}{ethanol wiping} \\
 &  & - & $2.66\pm0.05$\\ \hline 
 \multicolumn{4}{c}{}\\
 \multicolumn{4}{c}{}\\
 & \textbf{disc 3}& \textbf{disc 4a }&\textbf{disc 4b}\\
 \hline 
 start act. & $7.52\pm0.22$ & $16.3 \pm0.6$ & $6.59 \pm0.19$  \\ \hline
  \multirow{2}{*}{1st step}& strong HNO$_3$ & \multicolumn{2}{c}{ethanol bath}  \\
 & $4.91\pm0.25$& $12.2\pm0.4$ & $5.0\pm0.3$ \\ \hline
\end{tabular}
\caption{\label{tab:po210_results} $^{210}$Po activity in counts/min measured after having applied different cleaning procedures. In case of disc 2 and disc 4, both the front side (a) and the back side (b) of the sample have been measured.}
\end{table}

\begin{figure}[h]
  \begin{center}\includegraphics[
      width=1.0\columnwidth,
      keepaspectratio,
      angle=0]{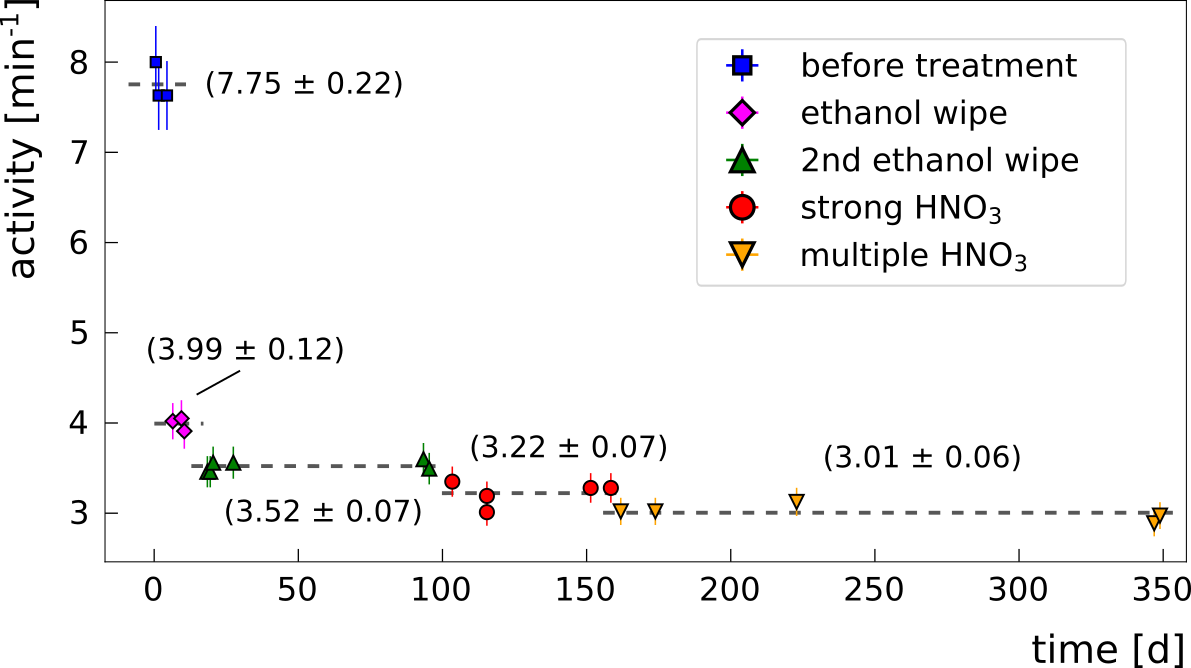}
    \caption{Evolution of the $^{210}$Po activity obtained after different cleaning steps for sample disc 2b. The x-axis gives the time distance of the single runs with respect to the first measurement of the sample.}\label{fig:disc2b_evol}
  \end{center}
\end{figure}

Table \ref{tab:po210_results} shows the measured $^{210}$Po activity for the four PTFE sample discs before and after the investigated procedures described in Table \ref{tab:c_proc}. In case of disc 1 and disc 2, a sequence of cleaning steps has been applied, whereas for discs 3 and disc 4 only a single cleaning procedure was tested. For disc 2 and disc 4, we measured the activity of both disc sides which are referred to as \textit{a} and \textit{b}, respectively. Since alphas cannot penetrate the PTFE disc, both sides are considered as independent samples. Fig. \ref{fig:disc2b_evol} is the visualization of the results obtained for disc 2b. For all samples, the best result was found after the first cleaning step when a $^{210}$Po reduction of up to 50\% was achieved in case of ethanol wiping (disc 2b). All subsequent cleaning steps did not show a further reduction of comparable size, independent of the applied procedure. This observation implies that half of the plated-out $^{210}$Po is located directly on the sample's surface and could be efficiently removed. The remaining contamination, on the other hand, was hardly accessible with our cleaning procedures. Since the multiple HNO$_3$ procedure did not show a further activity reduction, we can also exclude re-deposition of polonium on the disc surface from the cleaning liquid.

So far, we discussed only the removal of $^{210}$Po. However, the low energetic beta decay of $^{210}$Pb is also an issue for many low background experiments. Due to its chemical properties, the removal of lead achieved in the cleaning procedures is expected to be different from the obtained polonium results. Since the $\beta$-decay of $^{210}$Pb cannot be measured by means of alpha spectroscopy, we conclude on the lead removal only indirectly by analyzing the time evolution of the $^{210}$Po activity after each cleaning step. Due to the $^{210}$Po half-life of $138$\,d, this analysis requires a monitoring of the sample's polonium activity over a time of several weeks. After $\sim 50$\,d we can expect activity changes of 10\% at maximum, i.e., in the extreme cases when either all or no $^{210}$Pb has been removed. The required data is available for disc 1 and disc 2 for some of the cleaning procedures (see Fig. \ref{fig:disc2b_evol} for disc 2b). Our $^{210}$Po evolution model is based on the assumption that at the beginning of this study $^{210}$Pb was in a secular equilibrium with its daughter isotopes $^{210}$Bi and $^{210}$Po (see Section \ref{sec:1.1}). Thus, the measured $^{210}$Po activity before the first cleaning step, referred to as $A_0$, corresponds also to the activity of $^{210}$Pb. Due to different cleaning efficiencies, however, the relative amount of lead and polonium might change in the investigated procedures. After each cleaning step the time evolution of the $^{210}$Pb nuclide on the sample's surface is given by its radioactive decay
\begin{equation}
 N_{\text{Pb}}(t) = N_{\text{Pb}}^0 \cdot e^{-\lambda_{\text{Pb}}\cdot t} \approx N_{\text{Pb}}^0 = \text{const} \quad ,
\end{equation}
but is in good approximation constant until the next cleaning procedure is applied. 
The parameter $N_{\text{Pb}}^0$ thus refers to the number of atoms at the time right after the investigated cleaning step, i.e., at $t=0$ for the later modeled time evolution. The fit function for the $^{210}$Po evolution is then obtained from the differential equations describing the time evolution of the bismuth and polonium nuclides, respectively:
\begin{align}
 dN_{\text{Bi}}(t) / dt&= \lambda_{\text{Pb}}\cdot  N_{\text{Pb}}^0 - \lambda_{\text{Bi}} \cdot N_{\text{Bi}}(t)\\
 &\text{with}\quad N_{\text{Bi}}(t=0) = 
  \begin{cases}
    0 & \rightarrow \text{min. Bi}\\ \nonumber
    \frac{A_0}{\lambda_{\text{Bi}}} & \rightarrow \text{max. Bi}
  \end{cases} \\
 dN_{\text{Po}}(t) / dt & = \lambda_{\text{Bi}}\cdot  N_{\text{Bi}}(t) - \lambda_{\text{Po}} \cdot N_{\text{Po}}(t)\\
 &\text{with}\quad N_{\text{Po}}(t=0) = N_{\text{Po}}^0 \quad . \nonumber
 \label{eq:dg}
\end{align}
$^{210}$Bi is the direct mother isotope of $^{210}$Po and undergoes a beta-decay with a half-life of $5\,$days. Similarly to $^{210}$Pb, we cannot measure the bismuth reduction directly. In order to account for the impact of bismuth on the evolution of $^{210}$Po, we treat the two extreme scenarios for the $^{210}$Bi reduction separately. In the so-called \textit{min. Bi} scenario, the investigated cleaning step has removed all $^{210}$Bi from the sample disc. Thus, we assume the amount of bismuth atoms right after the cleaning step to be $0$. In the other extreme case, referred to as \textit{max. Bi}, the cleaning is assumed to haven't removed any bismuth and its amount at $t=0$ is given by the activity of $^{210}$Pb before the cleaning step. For all fits we used the most conservative start activity $A_0$ for \textit{max. Bi}, assuming previous cleaning procedures have removed neither $^{210}$Pb nor $^{210}$Bi. 
The start activity $A_0$ is measured for every sample disc before the beginning of our cleaning study (see Table \ref{tab:po210_results}). Therefore, the only fit parameters are $N_{\text{Pb}}^0$ and $N_{\text{Po}}^0$, respectively. 
We selected periods where we have monitored the $^{210}$Po activity for longer than $20$\,d after a cleaning step and determine the $^{210}$Pb activity during that period by fit. 
\begin{figure}[htbp]
  \begin{center}\includegraphics[
      width=0.7\columnwidth,
      keepaspectratio,
      angle=0]{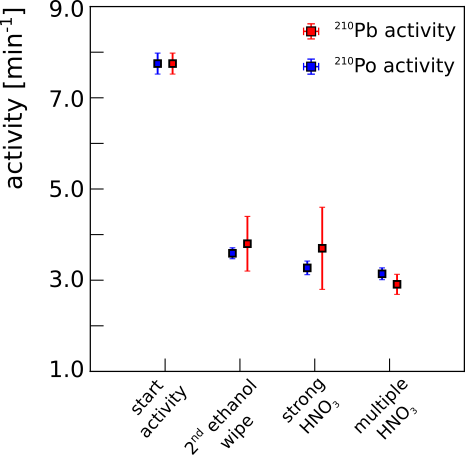}
    \caption{Fitted activities for $^{210}$Pb and $^{210}$Po after different cleaning steps applied to disc 2b. No difference in the cleaning efficiency between lead and polonium could be found.}\label{fig:pb_fit}
  \end{center}
\end{figure}
Fig. \ref{fig:pb_fit} shows the activities determined from $N_{\text{Pb}}^0$ (red) and $N_{\text{Po}}^0$ (blue) obtained for disc 2b after the different cleaning steps. For the error estimation we combined the fit results obtained assuming both extreme cases \textit{min. Bi} and \textit{max. Bi}, respectively. We could not find significantly different cleaning efficiencies between both radon daughters. In fact, the fitted reduction of $^{210}$Pb seems to follow the measured $^{210}$Po activity. Similar results have been obtained in the analysis of the polonium evolution of other sample discs.\\
\begin{table*}[t]
\centering
\renewcommand{\arraystretch}{1.5}
\smallskip
\begin{tabular}{c|c|l|c}
\textbf{chem. solution} & \textbf{sample}&  \multicolumn{1}{|c|}{\textbf{description of procedure}}& \textbf{red. factor}\\
 \cline{1-4}
 DI-water &1 &  DI-water bath for 1\,h incl. 15\,min US & $3.49\pm0.09$ \\ \cline{1-4}
 \multirow{6}{*}{ HNO$_3$ (5\%)} & 2 & acidic bath for  1\,h & $4.9\pm0.3$ \\ \cline{2-4}
 & 3 & acidic bath for  1\,h & $6.3\pm0.4$\\ \cline{2-4}
 & \multirow{2}{*}{4}& acidic bath for  1\,h & $11.5\pm0.4$ \\ 
 & & repeat procedure (2\,h in total)& $14.2\pm0.6$ \\ \cline{2-4}
 & 5 & acidic bath for  1\,h incl. 15\,min US & $13.7\pm1.4$ \\ \cline{2-4}
 & 6 & acidic bath for  2.5\,h incl. 15\,min US & $14.4\pm0.8$ \\\cline{1-4}
\multirow{2}{*}{ HNO$_3$ (32\%)} & \multirow{2}{*}{7} & acidic bath for  1\,h incl. 15\,min US & $23.2\pm2.5$ \\
& & repeat procedure (2\,h incl. 30\,min US in total) & $34\pm3$ \\ \cline{1-4}
\multirow{2}{*}{citric acid (11\%)} & \multirow{2}{*}{8} & acidic bath for  1\,h incl. 15\,min US & $13.2\pm0.6$ \\ 
& & repeat procedure (2\,h incl. 30\,min US in total) & $14.3\pm0.7$ \\ \cline{1-4}
acetic acid (1\%), &  \multirow{2}{*}{9} & 10\,min in acetic acid solution + & \multirow{2}{*}{$16.5\pm1.2$} \\ 
1\% HNO$_3$ + 3\% H$_2$O$_2$ & & 10\,min in (HNO$_3$ + H$_2$O$_2$) solution &  \\ 
\end{tabular}
\caption{\label{tab:pb212_reduct} Achieved $^{212}$Pb reduction for different cleaning chemicals. The acid concentrations are given by weight. Every treatment finished with DI-water rinsing. Besides the chemicals also other parameter as the usage of ultrasound (US) or the exposure time in the acid bath have been changed. In case of samples 4, 7 and 8 the procedure was repeated after measuring the $^{212}$Pb reduction for the first time. All reduction factors are determined with respect to the untreated sample.}
\end{table*}

\subsection{Removal of $^{220}$Rn daughters from PTFE surfaces}
\label{sec:1.3}
\begin{figure}[htbp]
  \begin{center}\includegraphics[
      width=0.95\columnwidth,
      keepaspectratio,
      angle=0]{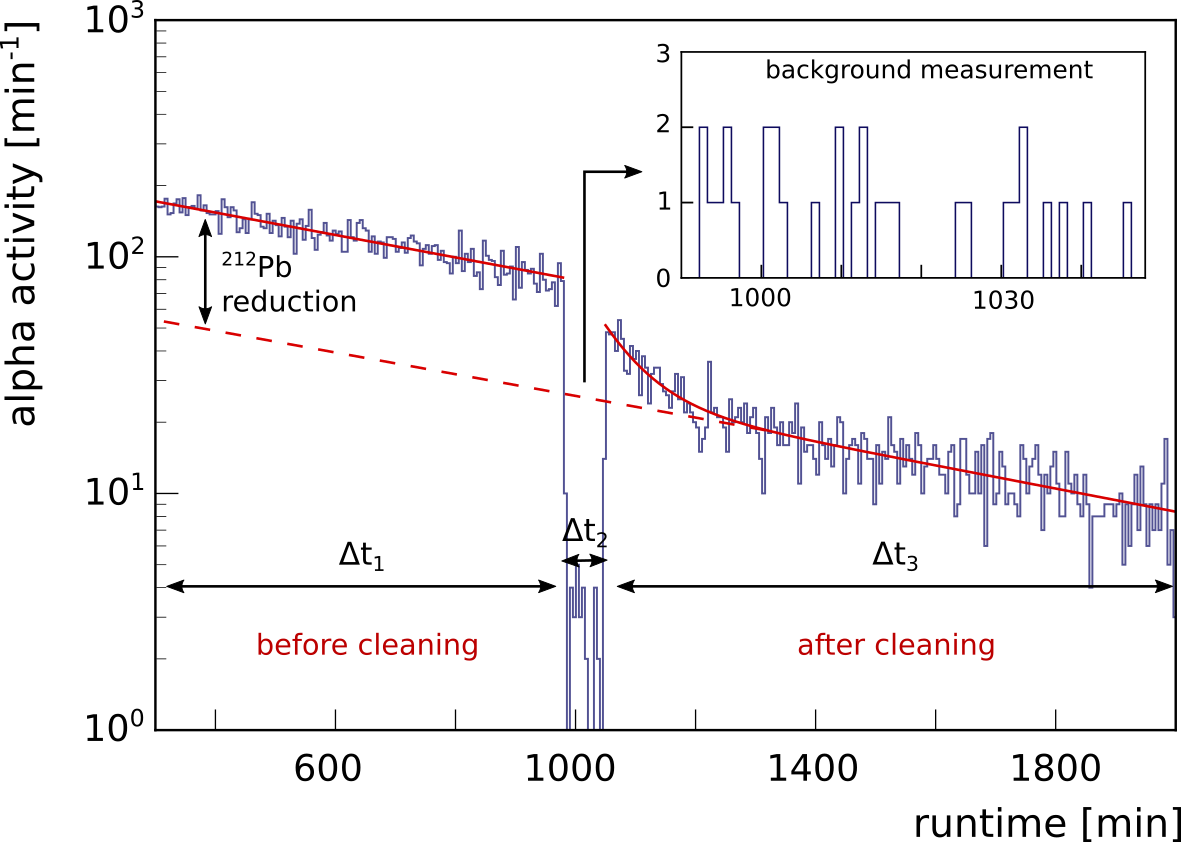}
    \caption{Exemplary $^{212}$Pb removal measurement (DI-water). At first the alpha activity of the sample disc is measured ($\Delta t_1$). During the cleaning process we monitor the background of the alpha-spectrometer ($\Delta t_2$ and inlet figure). After the cleaning, the now significantly reduced alpha activity is measured. ($\Delta t_3$).}\label{fig:pb212_cleaning}
  \end{center}
\end{figure}

In order to directly measure the removal of lead after cleaning, we performed further studies using PTFE samples loaded with $^{220}$Rn daughters. As depicted in Fig. \ref{fig:radon_chain} the decay chain of $^{220}$Rn includes only relatively short lived isotopes. When starting our measurements, only $^{212}$Pb with a half-life of $10.6$\,h and its daughter isotopes were present on the samples surface. As a consequence, the time evolution of the $\alpha$-decaying daughters $^{212}$Bi and $^{212}$Po is described by the radioactive decay of $^{212}$Pb. Thus, we can study the lead removal by means of alpha spectrometry on a much shorter time-scale with respect to $^{210}$Pb discussed in the previous section. For studying the $^{220}$Rn daughters, the samples have been newly loaded after each treatment.\\
Fig. \ref{fig:pb212_cleaning} shows an exemplary $^{212}$Pb removal measurement.
In our analysis, we combine the alpha-decays of both lead daughters $^{212}$Bi and $^{212}$Po, respectively. This can be done since the decays originate from two separate branches in the decay chain (see Fig. \ref{fig:decay-chain}) and are thus statistically independent. During the first period of the measurement ($\Delta t_1$), we monitored the alpha activity of the non treated sample disc after it has been loaded with radon daughters. From the number of counts, we can infer the alpha-activity at $t=0$ (the start time of $\Delta t_1$), which serves as a reference. The second phase ($\Delta t_2$) is a background measurement of the set-up. It was performed by replacing the sample with an unloaded sample disc. In all runs we found $^{220}$Rn daughters as background events. Our hypothesis is that lead atoms evaporate from the PTFE in the vacuum and plate-out on surrounding surfaces such as the PIN diode. The background measurement was done while in parallel the investigated cleaning procedure was applied. In the last phase of the run ($\Delta t_3$), the alpha activity of the cleaned sample is measured and the reduced activity is extrapolated to $t=0$. From comparison with the start activity before cleaning, the reduction factor is determined. The background of the detector is taken into account. Note that it does not get reduced by the cleaning but shows a time evolution given by the decay of $^{212}$Pb. Different cleaning efficiencies of $^{212}$Pb and $^{212}$Bi could impact the measurement shortly after cleaning. This effect is visible in Fig. \ref{fig:pb212_cleaning} (DI-water procedure) at the beginning of $\Delta t_3$ where it takes about $200\,$min until lead and bismuth reach an equilibrium.\\
We tested different chemicals such as nitric acid (HNO$_3$), citric acid (C$_6$H$_8$O$_7$) or hydrogen peroxide (H$_2$O$_2$) which are commonly used for cleaning detector materials \cite{cuore,exo_det_const,copper_clean}. In addition, we study the impact of other parameters on the cleaning efficiency such as an ultrasound treatment or the exposure time in the acidic bath. The results are summarized in Table \ref{tab:pb212_reduct}. Each sample was loaded only once with radon daughters before the measurement. In case of samples 4, 7 and 8 the investigated procedure was applied a second time after having determined the $^{212}$Pb reduction achieved in the first cleaning step. All reduction factors in Table \ref{tab:pb212_reduct} are determined with respect to the untreated sample.\\
A sample treated in DI-water with ultrasound, where a $^{212}$Pb reduction factor of 3.5 was found, serves as a reference. The 5\% (by mass) nitric acid treatment for 1\,h gave inconsistent results among the different samples but the averaged $^{212}$Pb reduction was about twice as efficient as the water procedure. A higher reduction could be achieved by increasing the exposure time in the acid or by using ultrasound during the treatment. A maximum reduction factor of $\sim 14$ was found for the 5\% nitric acid treatment which could not be further improved by a longer exposure. In case of the 32\% nitric acid, the $^{212}$Pb removal could be doubled, reaching a reduction factor of up to $(34\pm3)$ depending on the time in the acidic bath. The citric acid was found to be as efficient as the 5\% nitric acid solution. For the acetic acid procedure applied to sample 9, the PTFE disc was put for 5\,min into an acetic acid (1\%) solution followed by the  immersion for 5\,min into a HNO$_3$ (1\%) + H$_2$O$_2$ (3\%) solution. Thereafter, the acid baths were renewed and the procedure was repeated which results in a total exposure of 10\,min to both chemical solutions (acetic acid and HNO$_3$ + H$_2$O$_2$). The usage of acetic acid and hydrogen peroxide for lead removal has been studied, e.g., in \cite{lead_removal} for chromium surfaces. The reduction factor measured for our procedure was $(16.5\pm1.2)$, slightly higher than the reduction achieved for citric acid and nitric acid (5\%). For all tested procedures, however, the reduction of $^{220}$Rn daughters was found to be much larger with respect to the results obtained for $^{222}$Rn daughters.

\subsection{Comparison of $^{210}$Pb and $^{212}$Pb cleaning results}
\label{sec:1.4}
Since both lead isotopes are chemically identical, the reason for the different removal achievements is thought to be caused by the migration of radon daughters from the surface into the bulk material. We suppose that diffusion and the recoil energy induced by alpha-decays are possible transport mechanisms. In \cite{radon_diffusion}, the diffusion of $^{222}$Rn in polyethylene was determined by measuring the profile of $^{210}$Po inside the material. They obtained diffusion lengths at a millimeter scale. Also for our samples the diffusion of radon is one mechanism to transport long-lived radon daughters into the PTFE bulk. Since $^{222}$Rn features the longer half-life, it is expected to diffuse more efficiently into the PTFE than $^{220}$Rn before disintegration which could explain the observed difference between the two radon chains. At present, we are not aware of any measurements of the diffusion of heavy metals in PTFE. However, due to the much longer lifetime of $^{210}$Pb, even a small diffusion constant might be sufficient to let $^{210}$Pb migrate essentially deeper into the material with respect to $^{212}$Pb.\\
In the $^{220}$Rn decay-chain, the radon is followed by $^{216}$Po which exhibits a very short half-life of $150$\,ms. Consequently, it is mostly the $^{212}$Pb which plates-out on the sample disc during the loading process. This is in contrast to $^{210}$Pb where the decay-chain of $^{222}$Rn includes different long-lived polonium and lead isotopes which all have the chance to plate-out on the PTFE. Trapped on the surface, the short-lived radon daughters continue to decay. If the subsequent decays include alpha-decays, the recoiling daughter nuclei can get implanted into the bulk material of the PTFE sample. For the $7.8$\,MeV decay of $^{214}$Po, the recoil energy is $144$\,keV \cite{nucl_data} which results in an average implantation depth of $\sim 60\,$nm as simulated using the SRIM software package \cite{srim}.

\section{Impact on xenon purity}
\label{2}
\begin{figure}[h] 
  \begin{center}\includegraphics[
      width=1\columnwidth,
      keepaspectratio,
      angle=0]{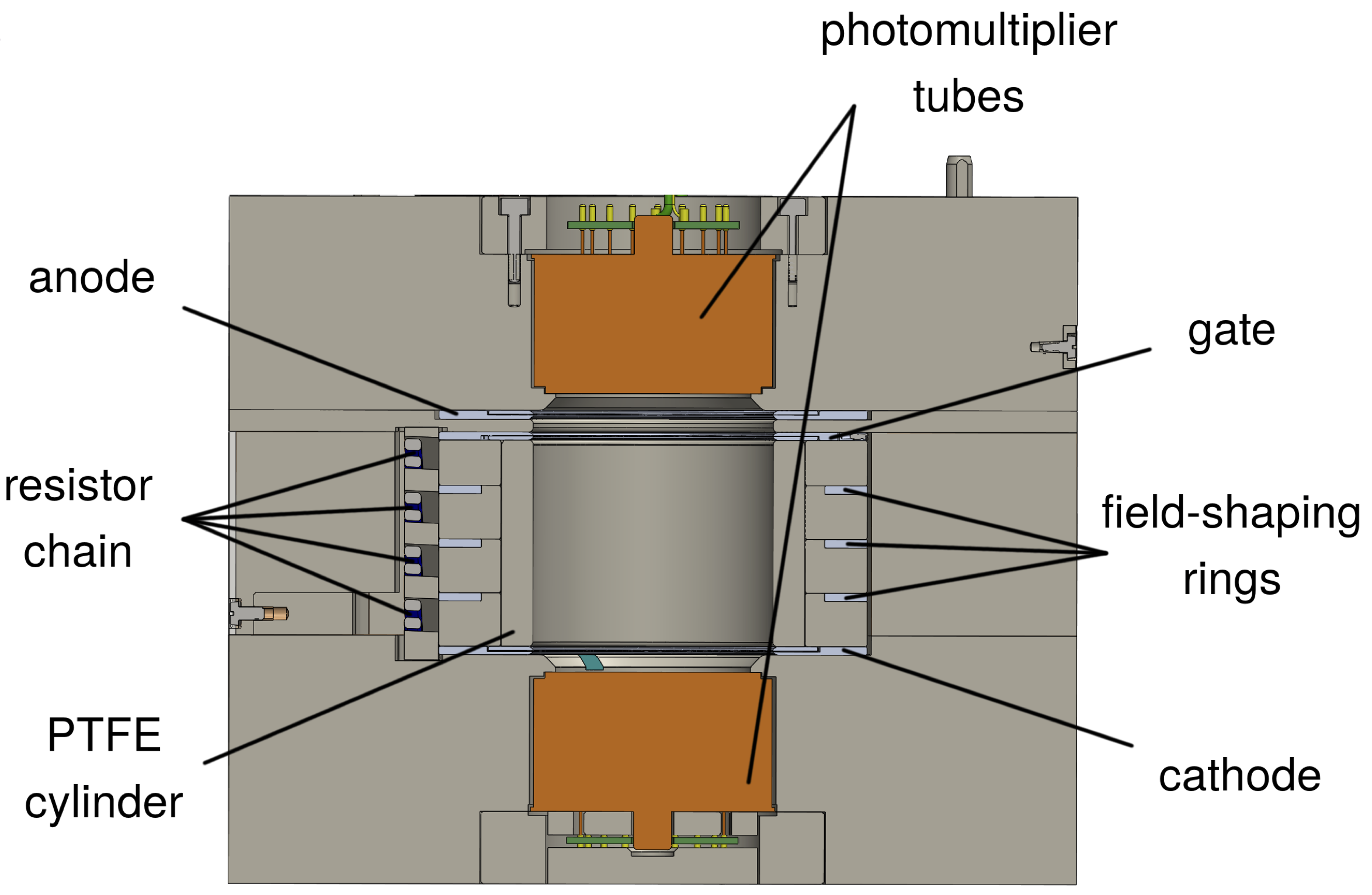}
    \caption{Lateral cut of HeXe TPC. Particle interactions with the xenon atoms create scintillation light signals that are detected by two photomultiplier tubes at the top and bottom of the active volume. In our study, we apply surface cleaning methods to the exchangeable sample PTFE cylinder, as detailed in the text.}\label{fig:TPC}
  \end{center}
\end{figure}

In addition to the ability of a cleaning procedure to remove radon daughters from PTFE, the potential effect from cleaning chemical residues on the detector performance needs to be considered. To study this effect, we employ the dual-phase TPC of the so-called HeXe set-up, sketched in Fig. \ref{fig:TPC} (also employed in \cite{PTFE_trans}.
The TPC contains liquid xenon and a layer of gaseous xenon on top. Particles interacting in the xenon create scintillation light and ionization electrons. The light signal is detected by two photomuliplier tubes at the top and bottom of the xenon volume, respectively. The electrons are drifted upwards by an electric field. At the liquid-gas interface they are extracted and create a second, delayed scintillation light signal. A PTFE cleaning recipe could potentially induce trace amounts of electronegative impurities in the xenon, which can hinder the charge collection via electron attachment. As shown in the previous sections, the 32$\%$ HNO$_{3}$ surface treatment shows a large removal factor for radon daughters (see Table \ref{tab:pb212_reduct}). Motivated by this, we tested the impact of 32$\%$ HNO$_{3}$ on the chemical purity of the xenon inside the TPC.
\subsection{Sample preparation and measurements}
\label{2.1}
The wall of the TPC's active volume has the shape of a hollow cylinder made of PTFE, as indicated in Fig. \ref{fig:TPC}. The height and diameter of the cylinder is $\sim\,5~$cm. As it is exchangeable, we employ copies of it as individual samples. Before each measurement, a new sample is treated with the cleaning chemicals under investigation and inserted into the TPC. Thus, its surface gets exposed directly to the liquid xenon and the potential desorption and outgassing of the cleaning agent into the cold medium can be tested. The sample cylinder is made of $150\,$g PTFE and has a surface of $220\,$cm$^2$, whereby the total TPC's PTFE mass and surface is larger by a factor of 39 and 19, respectively. The TPC's active volume contains $\sim 345\,$g of liquid xenon, with a total mass of  $\sim 2.3\,$kg hosted in the cryostat. During a measurement, a homogeneous drift field of $\sim 390\,$V/cm is applied over a maximal drift length of $5\,$cm.  \\
In this study, we compare samples that undergo a basic cleaning with the neutral detergent ELMA CLEAN 65 \cite{elma} (blank sample) to samples cleaned in addition with 32$\%$ HNO$_{3}$ (test sample). For the detergent cleaning, the sample is immersed in an ultrasonic bath, filled with a $3\,\%$ mixture of ELMA CLEAN 65 and DI-water for $15\,$min at $40\,^\circ$C. Afterwards, the sample is rinsed in DI-water. A test sample is additionally submerged in 32$\%$ HNO$_{3}$ for $2\,$hours and then rinsed again. After rinsing, all samples are dried with a N$_{2}$ flow of $200\,$sccm in a leak tight stainless steel (SS) vessel at an absolute pressure of $50\,$mbar. After $3\,$days, the sample is inserted into the HeXe TPC. During this operation, the sample is always bagged and kept under a N$_{2}$ atmosphere. Thus, a possible contamination of the PTFE by air is avoided. This is important, because oxygen, water vapor and other impurities do affect the detector performance \cite{aprile_doke}. After the sample insertion, the TPC is evacuated within $20\,$hours to a pressure of $\mathcal{O}(10^{-4}~\text{mbar})$. Subsequently, it is filled with liquid xenon and the measurement is started. During the measurement, the xenon target is permanently purified by looping xenon gas continuously through a hot gas purifier which is installed in the gas purification system connected to the TPC's cryostat. All of the described steps follow a strictly timed schedule and the system parameters, such as the recirculation flow, are adjusted to a comparable level throughout the measurements. The purity of the xenon is quantified by the so-called electron lifetime parameter $\tau$ \cite{aprile_doke}. It depends inversely on the concentration of the present impurities and the electric field-dependent electron attachment rates. As the exact composition of impurities is usually unknown, one commonly expresses the impurity level equivalent to an O$_{2}$ concentration in units of ppb \cite{bakale}. The continuous xenon purification causes an increase of $\tau$ over time, up to a plateau when the outgassing rate of the detector materials, including the PTFE surface treatments, and the impurity removal rate, are equivalent. For the measurement of the $\tau$ time evolution, we employ the internal calibration source $^{83m}$Kr that allows for a clear event selection due to its characteristic delayed coincidence signature, as done in \cite{kr_source}. The amount of initial electrons N$_{\text{e}}(0)$, generated in the mono-energetic energy depositions, is reduced along the depth of the TPC since the electrons are trapped by impurities. Hence, only a fraction $\text{N}_\text{e}\text{(t) = N}_{\text{e}}(0) \cdot \text{exp}~(-  \text{t}/ \tau)$ can be detected per TPC depth, depending on $\tau$ and the electron drift time t. By comparing the time evolution of $\tau$ between blank and test samples, the potential degradation of the xenon purity due to the 32$\%$ HNO$_{3}$ treatment can be tested.
\subsection{Purity results and discussion}
\label{2.2}
\begin{figure}
  \begin{center}\includegraphics[
      width=1\columnwidth,
      keepaspectratio,
      angle=0]{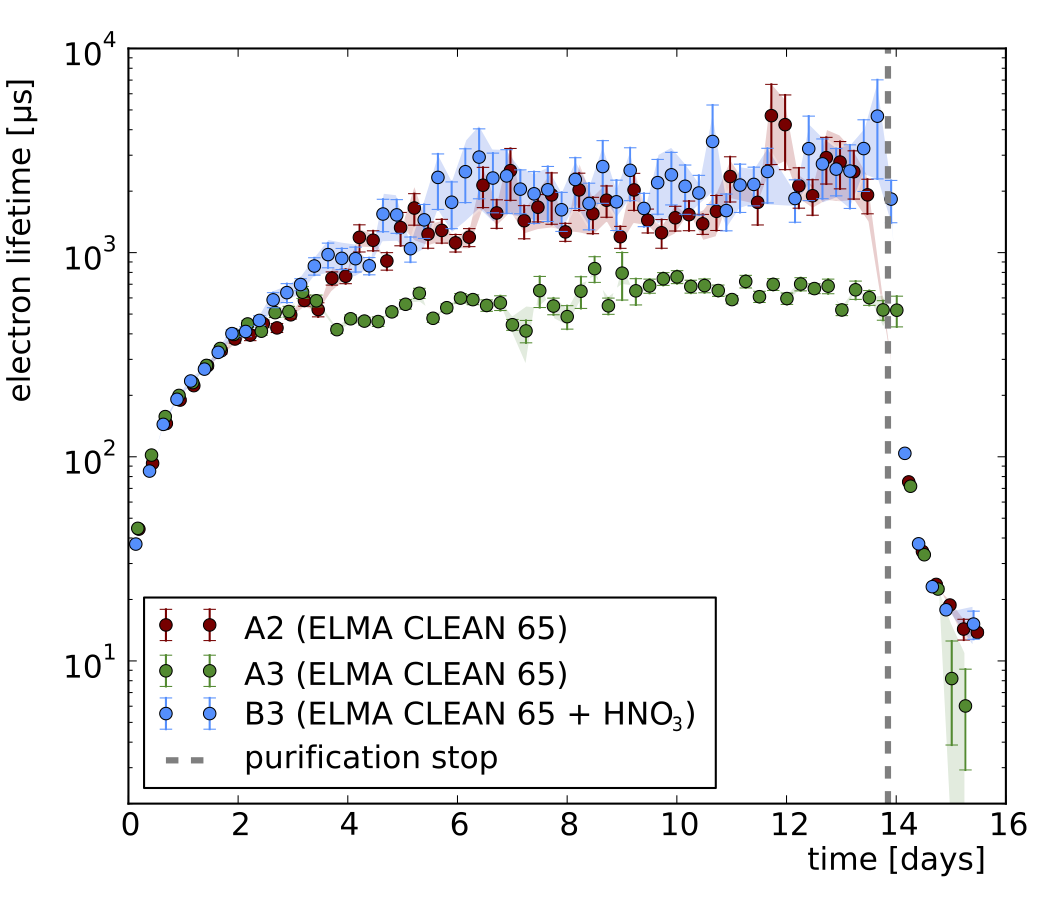}
    \caption{Electron lifetime evolution of blank samples A2, A3 and test sample B3. The error bars indicate the uncertainty from the electron lifetime fit and the shaded areas illustrate the average relative deviation between different analysis methods. After $\sim$ 14 days xenon purification is stopped, causing the electron lifetime to decrease.}\label{fig:electron_lifetime}
  \end{center}
\end{figure}

The $\tau$ evolution of the blank samples A2, A3 (ELMA CLEAN 65) and the test sample B3 (ELMA CLEAN 65 + HNO$_{3}$) are shown in Fig. \ref{fig:electron_lifetime}. 
The constant xenon purification causes an increase of $\tau$ in the first $\sim$ 14 days, after which the purification is stopped and consequently $\tau$ decreases. The error bars in Fig. \ref{fig:electron_lifetime} indicate the fit uncertainty and the shaded areas illustrate the average relative deviation between different analysis methods. The latter describes the effect of slightly altered data selection criteria and fitting procedures on the results. The A2 and B3 measurements show a comparable $\tau$ evolution which plateaus at $\sim 1700~\upmu$s. In this plateau region, the $\tau$ values exhibit large fluctuations due to a limited statistics and uncertainties in the analysis method. Therefore, we conservatively define the maximal achievable electron lifetime of the system when the average relative deviation among the analysis methods is larger than $15~\%$. Thus, we obtain a $\tau_{\text{max}} = 1000~\upmu$s that is achieved within the same purification time for both the sample treated with detergent only and the one cleaned with detergent and 32$\%$ HNO$_{3}$. In Fig. \ref{fig:electron_lifetime} we also show the evolution of A3, that serves as an example of the four further measurements, that all exhibit lower $\tau$ evolution, inconsistent among each other. The inconsistency is not fully understood. We hypothesize that slight variations of the set-up caused by the process of the PTFE sample cylinder insertion might yield to different xenon purification efficiencies. We convert the measured $\tau_{\text{max}}$ into an impurity concentration of $0.43$ ppb O$_{2}$ equivalent \cite{bakale} and determine the base outgassing rate of the entire system. For this purpose, we assume that the impurities are homogeneously distributed within the xenon and that the impurity removal efficiency of the gas purifier is 100$\%$. Furthermore, we presume that the xenon purification rate is equivalent to the impurity outgassing rate of the system at the measured impurity concentration. Thus, at a purification flow of $2.66~$slpm, we obtain an O$_{2}$ equivalent outgassing rate of $5\cdot 10^{-11}~(\text{mol~O}_{2})/\text{min}$. Presumably, the PTFE forming the structure of the TPC, contributes to this outgassing rate in large part. From the comparable measurements of A2 and B3, we conclude that the 32$\%$ HNO$_{3}$ treatment does not increase the impurity concentration above the base level of the system. Consequently, the xenon purity is not degraded within the system's sensitivity.

\section{Summary}
\label{summary}
In this paper, we investigated the potential of chemical surface treatments to remove radioactive radon daughters which originate from plate-out from ambient air. For metallic surfaces, such as copper or stainless steel, radon daughters can be successfully removed by chemical treatments where the surface gets etched on a micrometer-scale \cite{zuzel_wojcik}. In case of PTFE, however, the cleaning procedure does not remove the surface together with its contamination but the radon daughters are leached from the material.\\
In our study we measured a $\sim 50$\% reduction of the long-lived radon daughters $^{210}$Pb and $^{210}$Po on PTFE surfaces after different chemical treatments. Subsequent, additional cleaning steps on the same samples led only to a minor further reduction, independent on the used chemicals or procedure. Similar studies have been done with PTFE samples loaded with daughters from the $^{220}$Rn decay-chain. In contrast to $^{210}$Pb, the isotope $^{212}$Pb from the $^{220}$Rn chain was removed efficiently by the tested cleaning procedures. A reduction $>30$ could be achieved depending on the applied surface treatment.\\
In the second part of this work, we investigated if the treatment with 32\% HNO$_{3}$ affects the xenon purity in a liquid xenon TPC set-up. Therefore, it was applied to the PTFE surfaces directly in contact with the liquid xenon target. No additional impurity release above the system's base O$_{2}$ equivalent outgassing rate of $5\cdot 10^{-11}~(\text{mol~O}_{2})/\text{min}$ was measured within the sensitivity of the set-up. In further studies for large-scale detectors \cite{darwin}, TPCs with long drift lengths \cite{xenoscope} can be employed to reach higher sensitivities in combination with realistic xenon purification rates. The measurements presented in this work are the first indication that no degradation of the xenon purity is caused by a 32$\%$ HNO$_{3}$ surface treatment on PTFE.


\begin{thebibliography}{99}

\bibitem{strigari}
T. Marrod\'an Undagoitia and L. Rauch, \emph{Dark matter direct-detection experiments}, J. Phys. G43 no.1, 013001 (2016)

\bibitem{neutrino}
J. Schechter and J.W. F. Valle, \emph{Neutrinoless double-$\beta$ decay in SU (2) $\times$ U (1) theories}, Phys. Rev. D 25, 2951 (1982)

\bibitem{superCDMS}
SuperCDMS Collaboration, \emph{Projected Sensitivity of the SuperCDMS SNOLAB experiment}, Phys. Rev. D 95, 082002 (2017)

\bibitem{counterplateout}
J. Amsbaugh et al., \emph{An array of low-background $^3$He proportional counters for the Sudbury Neutrino Observatory}, Nucl. Inst. Meth. A 579, 1054 (2007)

\bibitem{borexinoplateout}
M. Leung, \emph{Surface Contamination From Radon Progeny}, AIP Conf. Proc. 785, 184 (2005)

 \bibitem{SR1paper}
E. Aprile et al. (XENON Collaboration), \emph{Dark Matter Search Results from a One Ton-Year Exposure of XENON1T}, Phys. Rev. Lett. 121, 111302 (2018)

\bibitem{LUXfullexposure}
 D.S. Akerib et al. (LUX Collaboration), \emph{Results from a search for dark matter in the complete LUX exposure}, Phys. Rev. Lett. 118, 021303 (2017)


\bibitem{pandaX}
L. Zhao et al. (PandaX Collaboration), \emph{PandaX: A deep underground dark matter search experiment in China using liquid xenon}, Mod. Phys. Lett. A33 (2018) no.30, 1830013 (2018)

\bibitem{LZ}
B.J. Mount et al. (LZ Collaboration), \emph{LUX-ZEPLIN (LZ) Technical Design Report}, arXiv:1703.09144 (2017)

\bibitem{xenon_instrument}
E. Aprile et al. (XENON Collaboration), \emph{The XENON1T Dark Matter Experiment}, Eur. Phys. J. C77:881 (2017)

\bibitem{darwin}
DARWIN Collaboration, \emph{DARWIN: towards the ultimate dark matter detector}, JCAP 1611 no.11, 017 (2016)

\bibitem{nexo}
nEXO Collaboration, \emph{nEXO Pre-Conceptual Design Report}, arXiv:1805.11142 (2018)


\bibitem{cuore}
C. Alduino et al, (CUORE Collaboration), \emph{CUORE-0 detector: design, construction and operation}, JINST 11 P07009 (2016)

\bibitem{gerda}
GERDA Collaboration, \emph{The GERDA experiment for the search of 0$\nu\beta\beta$ decay in $^{76}$Ge}, Eur. Phys. J. C 73 2330 (2013)

\bibitem{te_zou}
H. Zou et al., \emph{Quantifying the triboelectric series}, Nat. Commun. 10, 1427 (2019)

\bibitem{te_henniker}
J. Henniker, \emph{Triboelectricity in polymers}, Nature 196, 474 (1962)

\bibitem{te_liu}
C. Liu and A. J. Bard, \emph{Electrostatic electrochemistry at insulators}, Nat. Mater. 7, 505 (2008)

\bibitem{ion_fraction}
\emph{Neutralisation rate and the fraction of the positive $^{218}$Po-clusters in air}, Atmos. Environ. 37, 1057 (2003)

\bibitem{exo_ion}
EXO-200 Collaboration, \emph{Measurements of the ion fraction and mobility of alpha and beta decay products in liquid xenon using EXO-200}, Phys. Rev. C 92, 045504 (2015)

\bibitem{morrison:plateout_teflon}
E.S. Morrison et al., \emph{Radon Daughter Plate-out onto Teflon}, arXiv:1708.08534 (2017)

\bibitem{neutron_ptfe}
G.N. Vlaskin et al., \emph{Neutron Yield of the Reaction ($\alpha$, n) on Thick Targets Comprised of Light Elements}, At. Energy 117 5, 357 (2015)

\bibitem{rn220_source}
R. F. Lang et al., \emph{A Rn-220 source for the calibration of low-background experiments}, JINST 11, 04 (2012)

\bibitem{pablo}
P. Herrero G\'omez, \emph{Investigation of surface cleaning procedures for the removal of radon daughters from PTFE surfaces and their applicability in liquid xenon detectors}, Master Thesis, University of Heidelberg (2018)

\bibitem{crystal_ball}
T. Skwarnicki, \emph{Study of the radiative cascade transitions between the upislon-prime and upsilon resoncances}, PhD. thesis, Cracow Institute of Nuclear Physics, DESY (1986)

\bibitem{exo_det_const}
M. Auger et al. (EXO Collaboration), \emph{The EXO-200 detector, part I: detector design and construction}, JINST 7, 05 (2012)

\bibitem{zuzel_wojcik}
G. Zuzel and M. Wojcik, 
\emph{Removal of the long-lived $^{222}$Rn daughters from copper and stainless steel surfaces}, 
Nucl. Inst. Meth. A 676 140 (2012)

\bibitem{copper_clean}
E.W. Hoppe et al.,  \emph{Cleaning and passivation of copper surfaces to remove surface radioactivity and prevent oxide formation}, Nucl. Instr. and Meth. Phys. Res. A 579, 486-489, (2007)

\bibitem{lead_removal}
Kh. Gholivand et al., \emph{A novel surface cleaning method for chemical removal of fouling lead layer from chromium surfaces}, Applied Surface Science 256,7457-7461 (2010)

\bibitem{radon_diffusion}
W. Rau,  \emph{Meaurement of radon diffusion in polyethylene based on alpha detection}, Nucl. Instr. and Meth. Phys. Res. A 664, 65-70 (2012)

\bibitem{nucl_data}
IAEA Nuclear Data Services,
https://www-nds.iaea.org/

\bibitem{srim}
 J. F. Ziegler et al.
 \emph{SRIM - The stopping and range of ions in matter},
 Nucl. Instrum. Meth. B 268, 1818 (2010). http://www.srim.org/


 \bibitem{PTFE_trans}
 D. Cichon et al., \emph{Transmission of xenon scintillation light through PTFE}, JINST 15, 09 (2020)

 \bibitem{elma}
 Elma Ultrasonic, https://www.elma-ultrasonic.com

 \bibitem{aprile_doke}
E. Aprile and T. Doke, \emph{Liquid Xenon Detectors for Particle Physics and Astrophysics}, Rev. Mod. Phys. 82:2053-2097 (2010)
 
 \bibitem{bakale}
 G. Bakale et al., \emph{Effect of an Electric Field on Electron Attachment to SF6, N20, and O2 in Liquid Argon and Xenon}, J. Phys. Chem., 80 (23) (1976).
 
  \bibitem{kr_source}
A. Manalaysay et al., \emph{Spatially uniform calibration of a liquid xenon detector at low energies using 83m-Kr}, Rev. Sci. Instrum. 81:073303, (2010)
 
 
 
 \bibitem{xenoscope}
 \emph{Xenoscope set-up},\\
https://www.physik.uzh.ch/en/groups/baudis/\\Research/Xenoscope.html, University of Zürich (2020)




\end{thebibliography}
\end{document}